\documentclass{article}
\usepackage{spconf,amsmath,graphicx}
\usepackage{mathrsfs}
\usepackage{amsfonts}
\usepackage{amsthm}

\usepackage[ruled,linesnumbered]{algorithm2e}

\title{A Quantitative Metric for Privacy Leakage in Federated Learning }
\name{Yong Liu\sthanks{Work done as an intern at Ping An Technology (Shenzhen) Co., Ltd. }$^{ \dagger}$ \qquad Xinghua Zhu$^{\star}$\sthanks{These authors contributed equally to this work.} \qquad Jianzong Wang$^{\star}$\sthanks{Corresponding author: jzwang@188.com.}\qquad Jing Xiao$^{\star}$}

\address{$^{\star}$ Ping An Technology (Shenzhen) Co., Ltd. \qquad $^{*}$ National University of Singapore}
\begin{document}
%
\maketitle
%
\begin{abstract}
In the federated learning system, parameter gradients are shared among participants and the central modulator, while the original data never leave their protected source domain.
However, the gradient itself might carry enough information for precise inference of the original data.
By reporting their parameter gradients to the central server, client datasets are exposed to inference attacks from adversaries.
In this paper, we propose a quantitative metric based on mutual information for clients to evaluate the potential risk of information leakage in their gradients.
Mutual information has received increasing attention in the machine learning and data mining community over the past few years.
However, existing mutual information estimation methods cannot handle high-dimensional variables. 
In this paper, we propose a novel method to approximate the mutual information between the high-dimensional gradients and batched input data.
Experimental results show that the proposed metric reliably reflect the extent of information leakage in federated learning.
In addition, using the proposed metric, we investigate the influential factors of risk level.
It is proven that, the risk of information leakage is related to the status of the task model, as well as the inherent data distribution.
\end{abstract}
\section{Introduction}
\label{sec:intro}
In the contemporary AI industry, there is an ever-rising quest for organized data.
In most industries, though, large amounts of data exist in isolated devices and institutes, being wasted away under the restriction of security or privacy regulations.
Federated learning (FL) \cite{konevcny2016federated,konevcny2016federated_2,he2020fedsmart} is devised to activate the isolated data sources.
Distinguished from the centralized machine learning, in an FL system, data do not leave their protected source locations \cite{kong2020network}.
Instead, the model parameter gradients are reported to a central modulator for global model aggregation.

The FL framework is a promising ideology.
Yet, when it comes to application, many data holders still lack incentives to participate in the FL process \cite{zhu2020empirical,zhu2019federated}.
One of the major concerns lies within the verifiability information security.
It has been proved that by observing parameter gradients, an adversary can make precise inference on the raw input data \cite{zhu2019deep,geiping2020inverting}.
Although provable encryption schemes, such as homomorphic encryption \cite{rivest1978data,acar2018survey}, secure multi-party computation \cite{shamir1979share,goldreich1998secure,mohassel2018aby3} and secret sharing, has been proposed to guarantee information security, their implementation and operation are to costly for practical applications.
In terms of data obfuscation techniques, such as differential privacy \cite{dwork2006calibrating,dwork2014algorithmic}, engineers need to empirically balance the privacy level and federated model performance.
From a data holder's perspective, the level of security provided by such designs is too arbitrary to be convincing.
Therefore, a quantifiable and universal metric is essential to promote incentives for data contribution in the FL systems.

Some may advocate the added noise level in the differential privacy scheme as an indicator of security degree.
But there is no proof of a quantifiable relationship between the noise level and the information leakage risk.
It still remains an open problem to systematically define the sufficient noise level for a differentially private model \cite{Lee2011}.

A practical metric for information leakage risk should satisfy the following properties:
\begin{itemize}
\item {\it Scale invariance} - the quantitative value should have the same meanings under different circumstances.
\item {\it Interpretability} - the metric should be inline with provable information bounds.
\item {\it Model-agnostic} - the model itself can be distributed as a blackbox to clients.
\end{itemize}
In the information theory, mutual information (MI) is a measure of the common information between two random variables.
It provides a theoretically provable, universal and quantifiable metric for the amount of information leakage on one variable given the other.
In the machine learning society, studies have been dedicated to estimate the MI between observable variables (model parameters, gradients, etc.) and the original data.
However, existing methods were mostly based on discrete variables, or made risky assumptions about the probability density functions \cite{belghazi2018mine, noshad2019scalable}.

\begin{figure}[htbp]

\centering
  \includegraphics[width=0.5\textwidth]{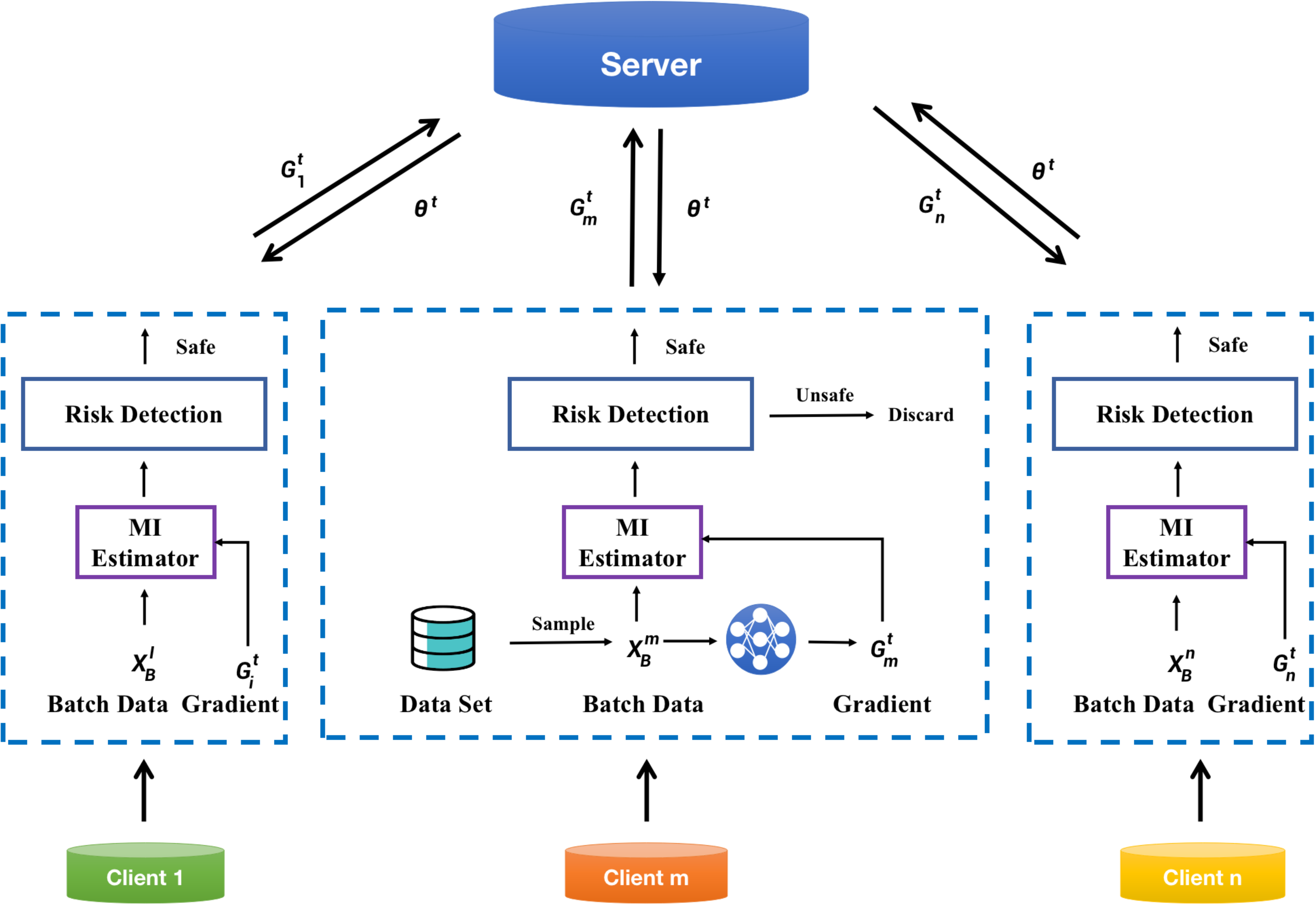}
%
\caption{Structure of an FL system with risk pre-alarm.}
\label{fig:fl}
\end{figure}

For the FL framework, we aim to measure the risk of information leakage before reporting the calculated model parameters to the central server.
Consequently, a client can make informed decision on whether it is safe to upload its model parameters. 
Therefore, the corresponding MI input / output variables are batched raw data and the computed gradients, respectively.
Both variables easily have hundreds of dimensions, well beyond the scope of discussion in previous works.
In this paper, we propose a novel hierarchical mutual information estimation method, H-MINE, for high-dimensional MI approximation.
The proposed method is then applied to estimate the risk of information leakage in an FL client under various experimental settings.
The main contributions of this paper are as follows: 
\begin{itemize}
\item Propose a novel hierarchical model, H-MINE, for robust and efficient high-dimensional MI estimation.
\item Apply H-MINE in the quantification of information leakage risk in FL systems.
\item Verify the credibility of H-MINE through comparison of inference attack results.
\item Analyse the inherent influential factors of information security in FL systems. 
\end{itemize}

\section{Proposed Method}
\label{sec:pagestyle}

\subsection{Risk Pre-Alarm in FL Systems}

\newcommand{\params}{\mathbf{\theta}}

In the federated stochastic gradient descent (FedSGD) algorithm, clients and the central server communicates iteratively to jointly optimize the global task model $f_\params$ \cite{konevcny2016federated}.
In a communication round $t$, a client $i$ obtains the current global model parameters $\params^t$ from the central server.
Client $i$ calculates the parameter gradient $G_i^t = \nabla_\params f_\params^t(X_B^i)$ with a batch sampled from local dataset $\mathcal{D}_i$.
That is, the batched data $X_B^i = \{x_1^i, x_2^i, ..., x_B^i\} \subseteq \mathcal{D}_i$, where $B$ is the batch size.
Client gradients $G_i^t$, $i=1,...,N$, are sent to the central server.
On the central server, $G_i^t$'s are aggregated to update the global model, such that $\params^{t+1} \gets \params^t - \eta \sum_i G_i^t$, where $\eta$ is the learning rate.

In this paper, we assume a modest security environment, where all participants are honest-but-curious.
Both the clients and central server would not try to poison the learning process, but may probe the underlying raw data when they have access to other participants' public information $G_i^t$.
Therefore, clients are susceptible to inference attacks from the central server or other intercepting adversaries.
We propose to implement an information leakage risk estimator on the client side, so that a client can be alarmed of the potential risk before publishing its gradient information (see Fig.\ref{fig:fl}).
In this paper, the information leakage risk is quantified by the mutual information between the batched raw data $X_B^i$ and the gradient $G_i^t$, i.e., $I(X_B^i; G_i^t)$.

\subsection{The Neural Estimator for Mutual Information}

Without loss of generality, the datapoints in a client dataset are assumed to be independently and identically distributed (IID), i.e., $x_j^i \sim \mathcal{X}^i$, $j=1,...,|\mathcal{D}_i|$.
It follows that $G_i^t(X_B^i) \sim \mathcal{G}_i^t$.

As discussed in previous sections, MI estimation is non-trivial, especially for high-dimensional random variables.
Belghazi et al. proposed to solve this problem with a neural network \cite{belghazi2018mine}.
In fact, the MI between two random variables $I(X; G)$ is equivalent to the Kullback-Leibler Divergence $D_\text{KL}[\mathbb{P}_{XG}||\mathbb{P}_{X} \otimes \mathbb{P}_{G}]$.
The Donsker-Varadhan representation \cite{DV} of the KL-Divergence gives a lower bound on $I(X_B; G^t)$:
\begin{align*}
    I(X_B;G^t)&=D_{KL}[\mathbb{P}_{X_B G^t}||\mathbb{P}_{X_B} \otimes \mathbb{P}_{G^t}] \\
     & \geq \sup_{T \in \mathcal{T}}\mathbb{E}_{\mathbb{P}_{X_B G^t}}[T] - \log (\mathbb{E}_{\mathbb{P}_{X_B} \otimes \mathbb{P}_{G^t}}[e^{T}]),
\end{align*}
where $\mathcal{T}$ can be any class of functions $T:(\mathcal{X}^i, \mathcal{G}_i^t) \rightarrow \mathbb{R}$ satisfying the integrability constraints of the Donsker-Varadhan theorem.
As proposed by Belghazi et al, choosing a neural network as $\mathcal{T}$ transforms the MI estimation problem into a network optimization one.
This transformation exploits the flexibility of neural networks in approximating arbitrarily complex functions,
but also benefits from the well-developed tools for network optimization.
The structure of the statistic network $T_\phi$ can be designed to accommodate different data types.
Its parameters $\phi$ are optimized with iterative sampling from the joint and marginal distributions of the $X_B$ and $G^t$.

\subsection{Hierarchical Mutual Information Neural Estimation (H-MINE)}

\begin{figure}[htbp]

\centering
  \includegraphics[width=0.5\textwidth]{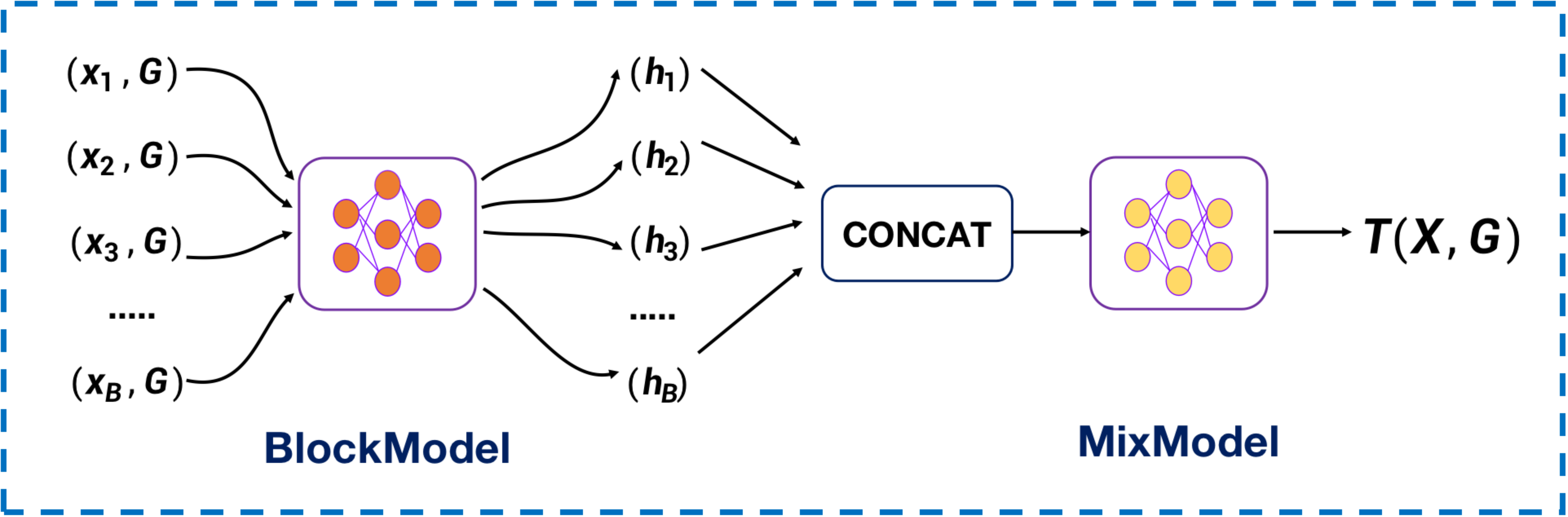}
%
\caption{H-MINE: the hierarchical statistic network.}
\label{fig:h-mine}
\end{figure}

Numerous architectures of the statistic network have been experimented in the literature.
They performed reasonably well as a constraining factor in tasks like generative adversarial networks \cite{doan2019line}, representation learning \cite{wen2020mutual}, and so on.
However, when we look into the accuracy of the estimated MI itself, the previously proposed statistic networks were not so successful.
When the variable dimensions increased, conventional statistic networks failed to converge or converged at insignificant values.
In this section, a novel statistic network structure is proposed that handles the high-dimensional input and output variables in the FL clients.

The structure of the proposed statistic network is illustrated in Fig.\ref{fig:h-mine}.
Particularly, we would like to make use of the fact that the data points $x\in \mathcal{D}_i$ are IID.
That is to say, the input variable in our statistic network, $X_B$, can be divided into $B$ independent random variables.
In our proposed statistic network, $x_m$ of batch $X_B$ and the gradient $G^t$ are grouped into a block $O_m = \{x_m,G^t\}$, $m = 1,...,B$.
Since $O_m$'s are identically distributed, a shared sub-network, called BlockModel, is used to map them into $B$ embedding vectors $h_m$.
The embedding vectors are then concatenated as input to the consequent MixModel, which outputs a single scalar in the final layer.
This statistic network structure is termed hierarchical MINE (H-MINE) for brevity.

As opposed to increasing the depth of the statistic network, H-MINE utilizes a shared sub-network to extract information from IID components.
The hierarchical structure of the proposed statistic network effectively reduces the dimensionality of the input variables.
The number of parameters in the input layer is also reduced by $B$ times, compared to a naive fully connected layer. The thorough procedure for MI approximation using H-MINE is elaborated in Algorithm \ref{algo:H-MINE}.

\SetKwInput{KwIn}{Inputs}
\SetKwInput{KwInit}{Initialize}
\begin{algorithm}[htb]
  \caption{H-MINE Algorithm.}
    \label{algo:H-MINE}
  \SetAlgoLined
  \KwIn{Task model batch size $B$, sample size $S$, task model status $\params^t$.}
  \KwInit{H-MINE parameters $\phi$.}
  \While{not converged}{
    \For{$k = 1,..., S$}{
    Generate batch samples $X_{B},\hat{X}_{B}$:\\
            $X_B \gets [x_1, ..., x_B]$, $\hat{X}_B \gets [\hat{x}_1, ..., \hat{x}_B]$ \\
        
    Calculate corresponding gradients:\\
    $G = \nabla_\params f_{\params^t}(X_B)$,
    $\hat{G} = \nabla_\params f_{\params^t}(\hat{X}_B)$ \\
    \For{$j = 1,..., B$}{
    $h_{j} = \textbf{BlockModel}_\phi(x_{j},G)$ \\
    $\hat{h}_{j} = \textbf{BlockModel}_\phi(x_{j},\hat{G})$ \\
    }
    $H = \{h_{1}, h_{2}, ... , h_{B}\}$,
    $\hat{H} = \{\hat{h}_{1}, \hat{h}_{2}, ... , \hat{h}_{B}\}$ \\
    $v_{k} = T_\phi(X_{B},G) = \textbf{MixModel}_\phi(H,G)$, \\
    $\hat{v}_{k} = T_\phi(X_{B},\hat{G}) =  \textbf{MixModel}_\phi(\hat{H},\hat{G})$\\
    }
    Evaluate the lower-bound: \\
    $V(\phi) =  \frac{1}{S}\sum_{k=1}^{S}v_{k}-\log(\frac{1}{S}\sum_{k=1}^{S}e^{\hat{v}_{k}})$ \\
    Update H-MINE parameters:\\
    $\phi \gets \phi + \nabla_{\phi}V(\phi)$ \;
    }
\end{algorithm}

\section{Results}

\subsection{Experiment Settings}

In this section, we evaluate the performance of our proposed methods with the Adult dataset \cite{Dua:2019}.
The task of the Adult dataset is to predict whether an individual's annual income exceeds \$50K.
It contains 14 private attributes about each individual, including education level, age, gender, occupation, and so on. 
The dataset contains 32,560 samples (7,841 positive and 24,719 negative).
Missing values are replaced with the medians in the dataset.
A logistic regression is used as the task classification model.

In H-MINE, the BlockModel is a multi-layer perception (MLP), which contains 3 fully connected layers with 200, 200, 5 neurons, respectively. 
The MixModel is a two-layer MLP with 500, 1 neurons, respectively. 
We use Adam \cite{kingma2014adam} optimizer to train H-MINE and learning rate $\alpha_{\phi} = 5 \times 10^{-5}$.

As a baseline of correlation metrics, the sum of all elements of covariance matrix $C(X_B, G^t) = \text{sum}\left(\text{cov}(X_B, G^t)\right)$ is also included in the subsequent figures.


\subsection{Convergence Analysis}

\begin{figure}[tbp]
    \centering
\begin{minipage}[b]{.49\linewidth}
  \centering
  \centerline{\includegraphics[width=\textwidth]{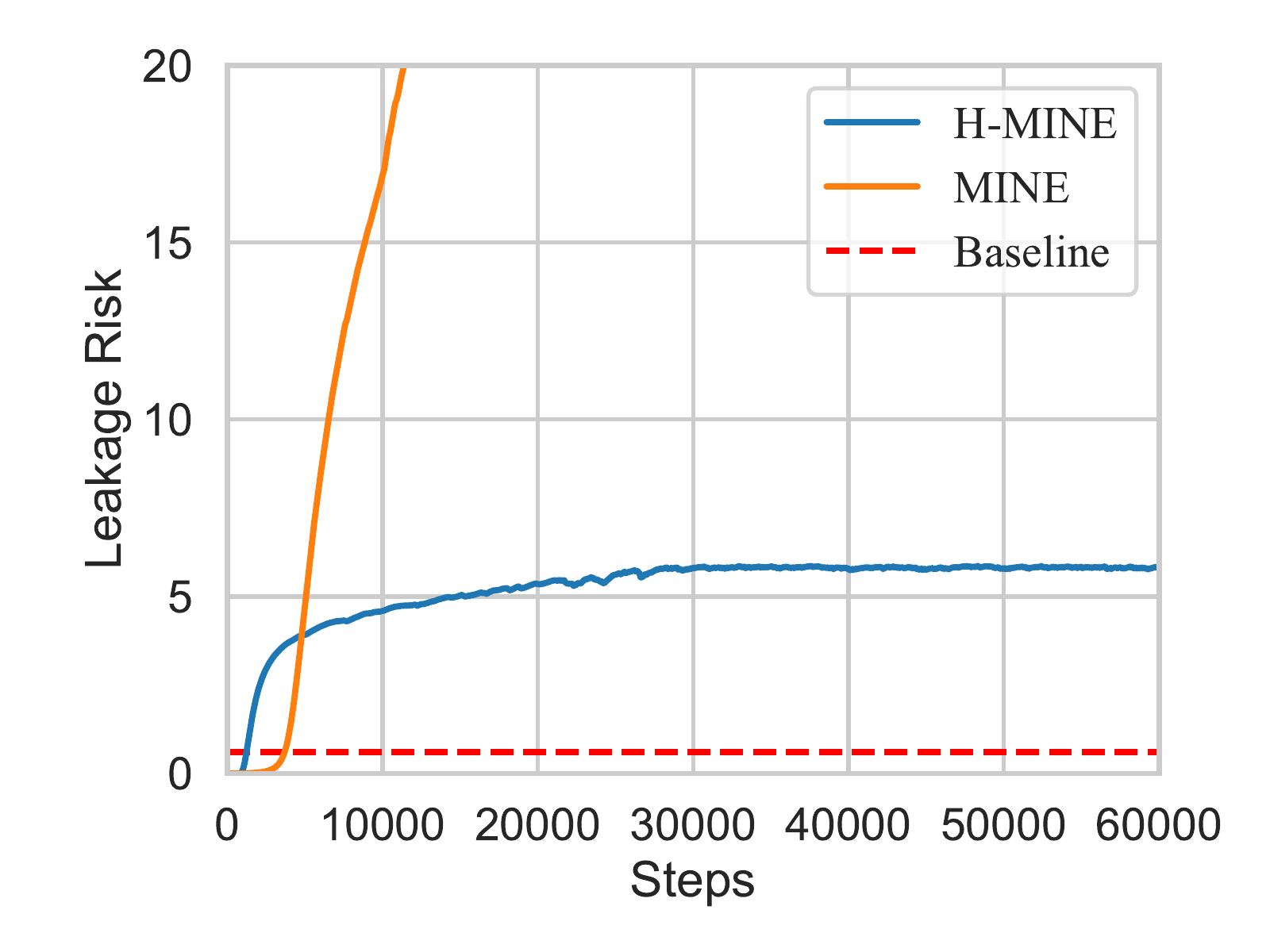}}
\end{minipage}
\begin{minipage}[b]{.49\linewidth}
  \centering
 \centerline{\includegraphics[width=\textwidth]{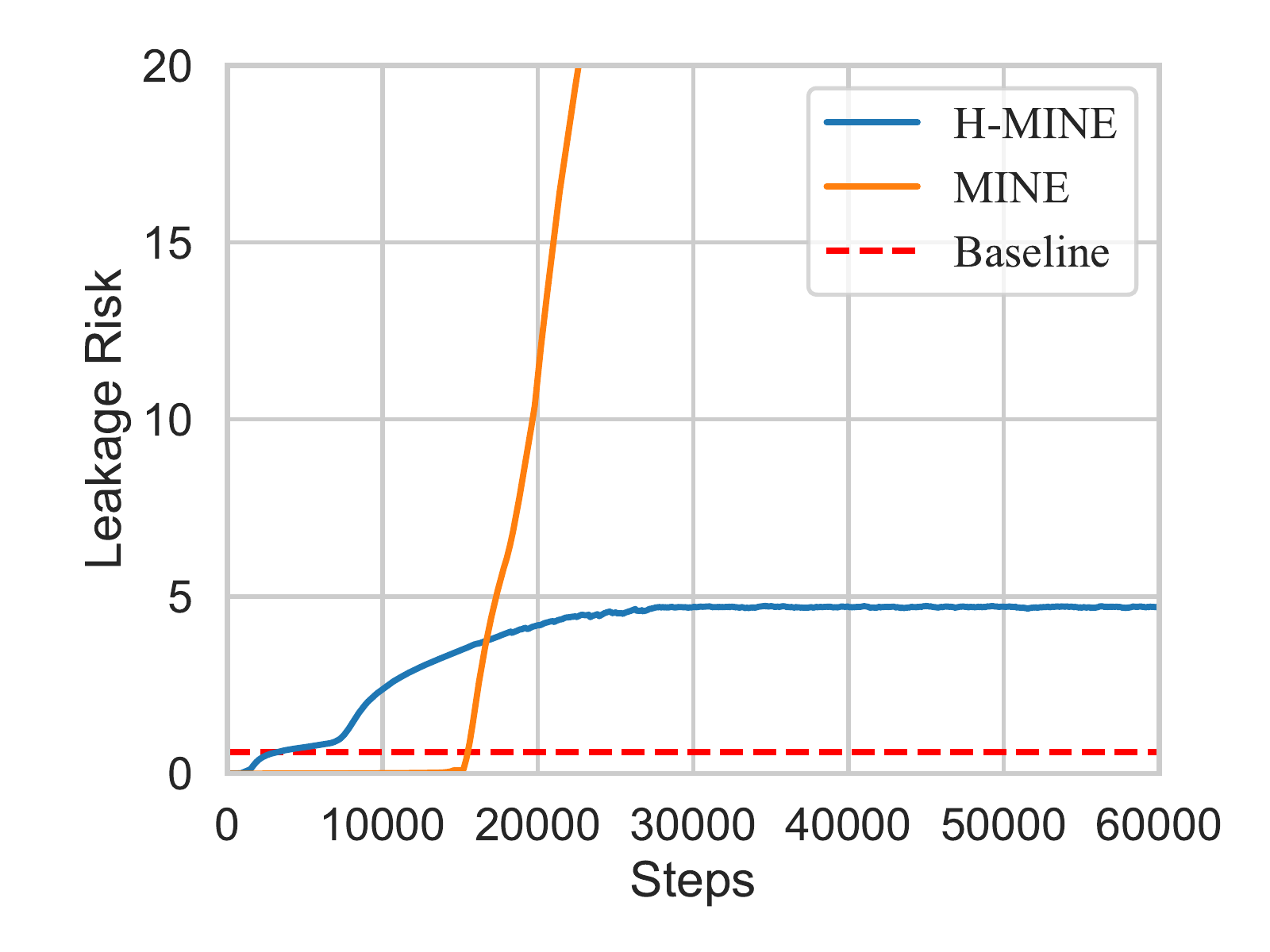}}
\end{minipage}
    \caption{Comparison of convergence speed of H-MINE and a conventional statistic network. Gradients are evaluated at $epoch=1$. Batch sizes are 1 and 3 respectively.}
    \label{fig:converge}
\end{figure}

In our proposed method, the MI is approximated via optimization of a statistic network.
The convergence curves of the proposed statistic network, H-MINE, are presented in Fig.\ref{fig:converge}.
The curves of a conventional statistic network, comprising of a three layer MLP \cite{belghazi2018mine}, are depicted for comparison.
In these experiments, the number of dimensions in the input variable multiplies as the batch size $B$ increases.
The conventional statistic network diverges in all attempted configurations, for its inability to handle high-dimensional variables.
In contrast, H-MINE conveges steadily even with increased batch size.






\subsection{Validation Via Inference Attack}

In this section, we use the DeepLeakage model to validate the accuracy of our estimated mutual information $I(X_B;G^t)$.
In theories, smaller mutual information between the two variables leads to increased difficulty in the inference attack.
Therefore, in the following experiments, we compare the inference error of the recovered data with our estimated $I(X_B;G^t)$ in various circumstances, and verify their correlation.

\begin{figure}[htbp]
    \centering
    \begin{minipage}[b]{.48\linewidth}
  \centering
  \centerline{\includegraphics[width=\textwidth]{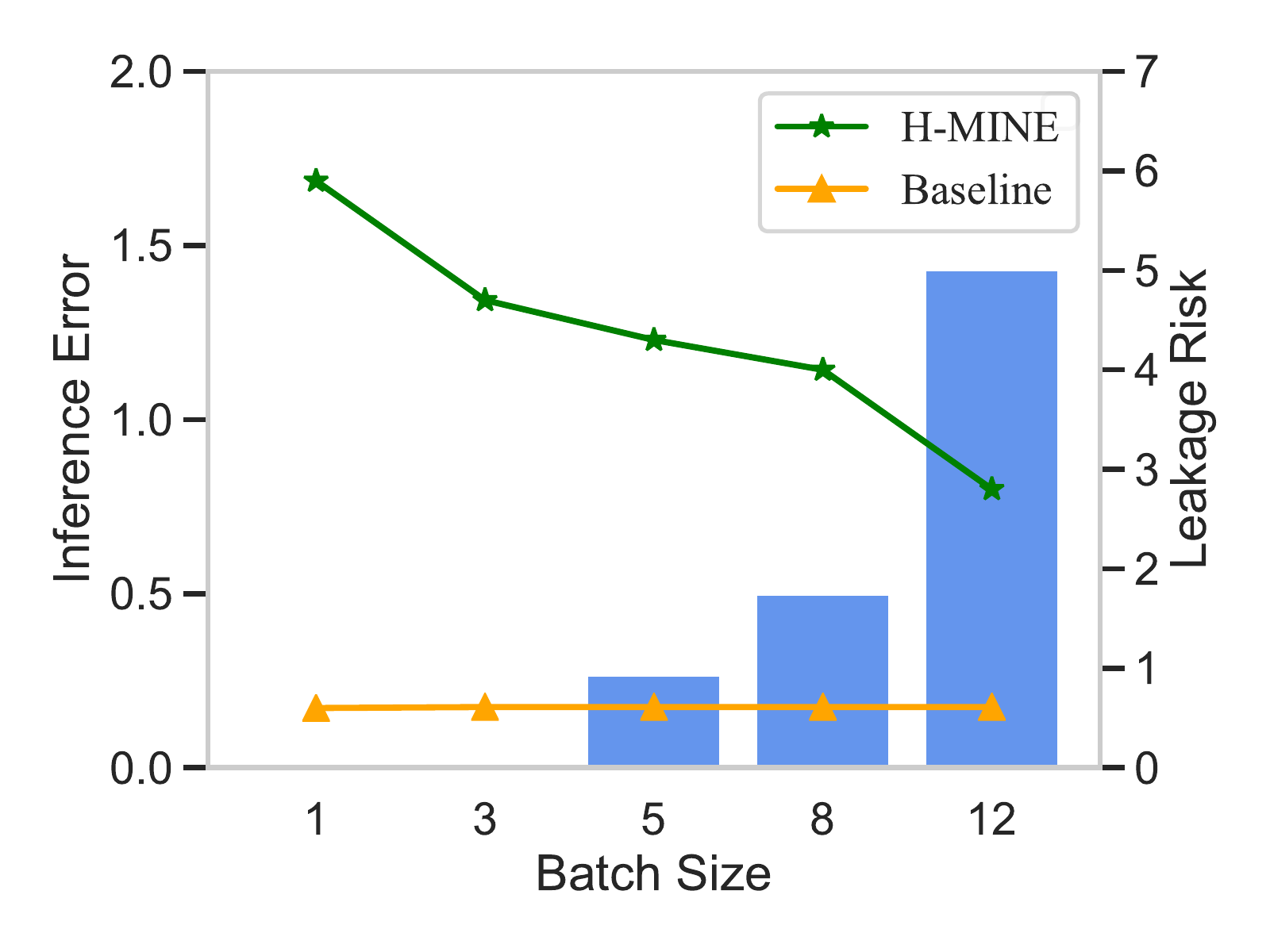}}
\end{minipage}
\begin{minipage}[b]{.48\linewidth}
  \centering
  \centerline{\includegraphics[width=\textwidth]{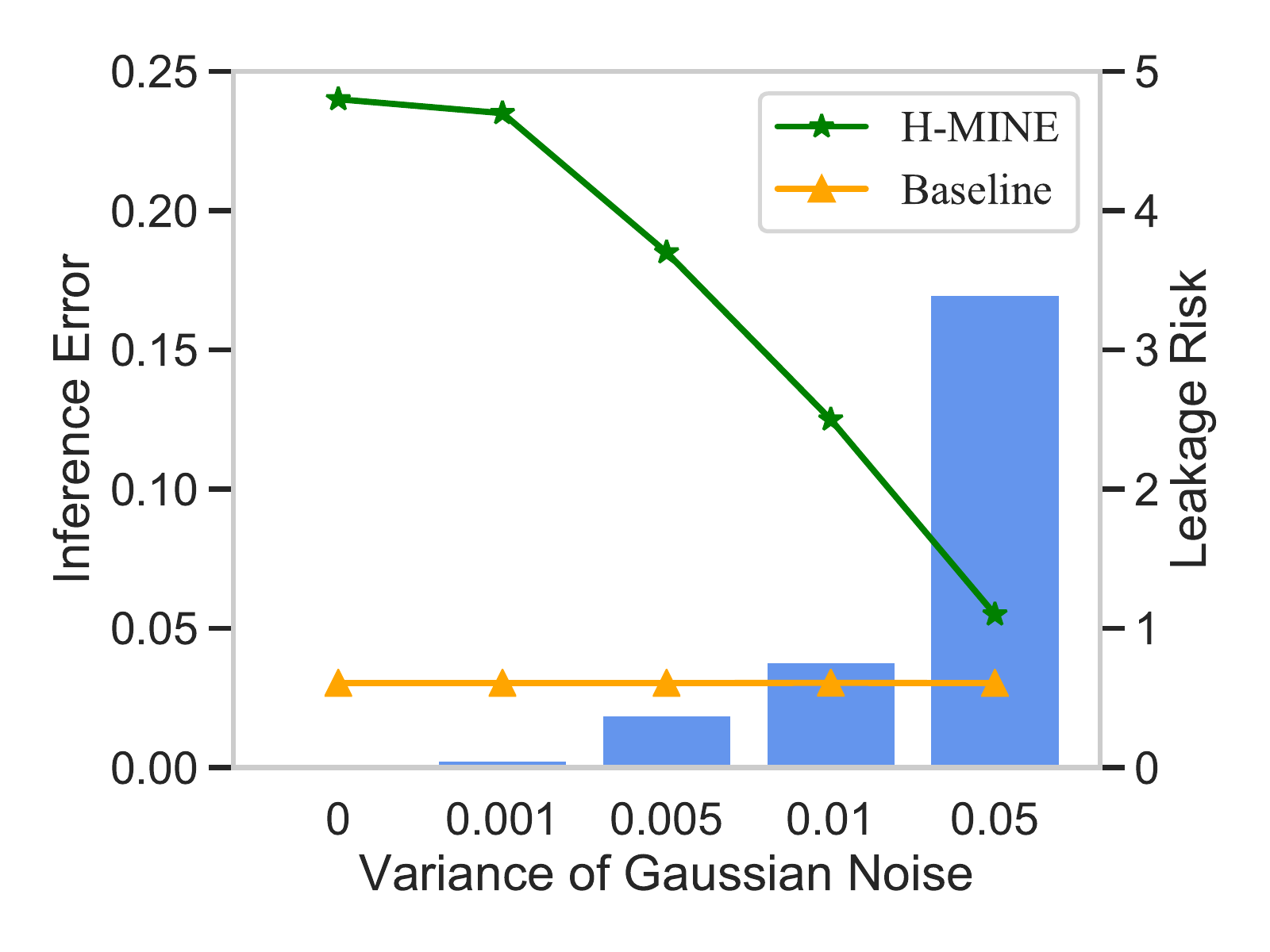}}
\end{minipage}
    \caption{Validation with DeepLeakage. Gradients are evaluated at $epoch=3$. For the added noise experiment, $B=3$.}
    \label{fig:deep_leakage}
\end{figure}

The DeepLeakage model in our experiments is a three-layer MLP that contains 100, 100, and 2 neurons, respectively.
For each configuration, the DeepLeakage is performed 5 times with random initialization, producing inference results $\hat{X}_B^{(k)}$, $k=1,...,5$.
The inference error of DeepLeakage is defined as $\epsilon = \text{var}(\hat{X}_B^{(k)})$.



Zhu et al. suggested two means to defend against inference attacks, namely, increasing the batch size, and adding noise to the gradients.
Fig.\ref{fig:deep_leakage} demonstrates the change of $I(X_B;G^t)$ and $\epsilon$ verses batch size (left) and added noise on the gradients (right).
As it shows, the estimated MI decreases as the batch size and noise level increase.
At the same time, the inference error increases quadratically.
The results are inline with our hypothesis, but also consistent with the results in \cite{zhu2019deep}.
In other words, our estimated MI can faithfully reflect the risk of inference attacks.



\subsection{Inherent Factors Affecting Information Leakage}

In this section, we analyze the factors affecting the risk of information leakage that are inherent in the data or the training process themselves.

\begin{figure}[htbp]
    \centering
\begin{minipage}[b]{.49\linewidth}
  \centering
  \includegraphics[width=\textwidth]{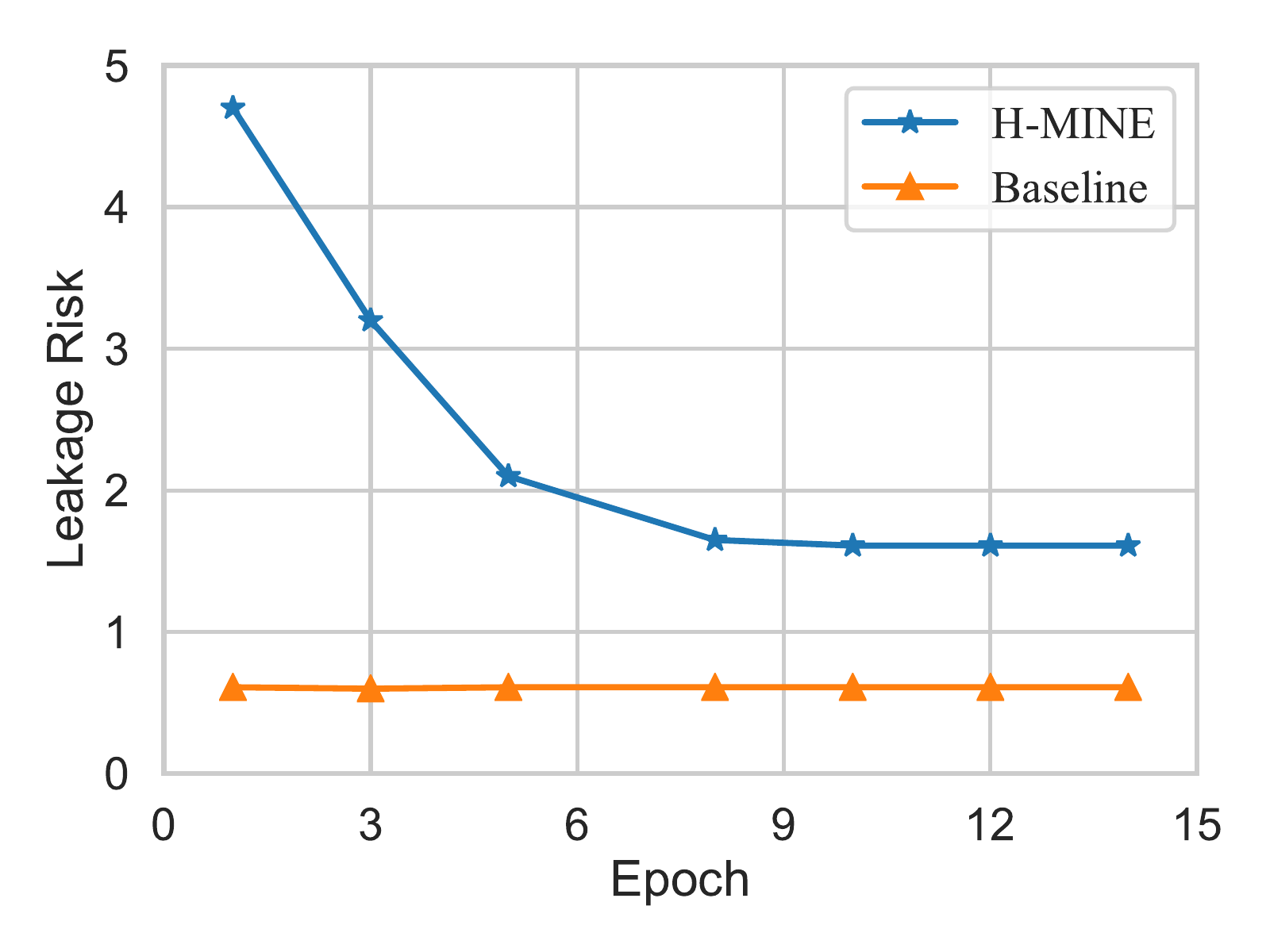}
  {\small (a) Leakage risk at different timesteps, batch size $B=3$.}
\end{minipage}
\begin{minipage}[b]{.49\linewidth}
  \centering
  \includegraphics[width=\textwidth]{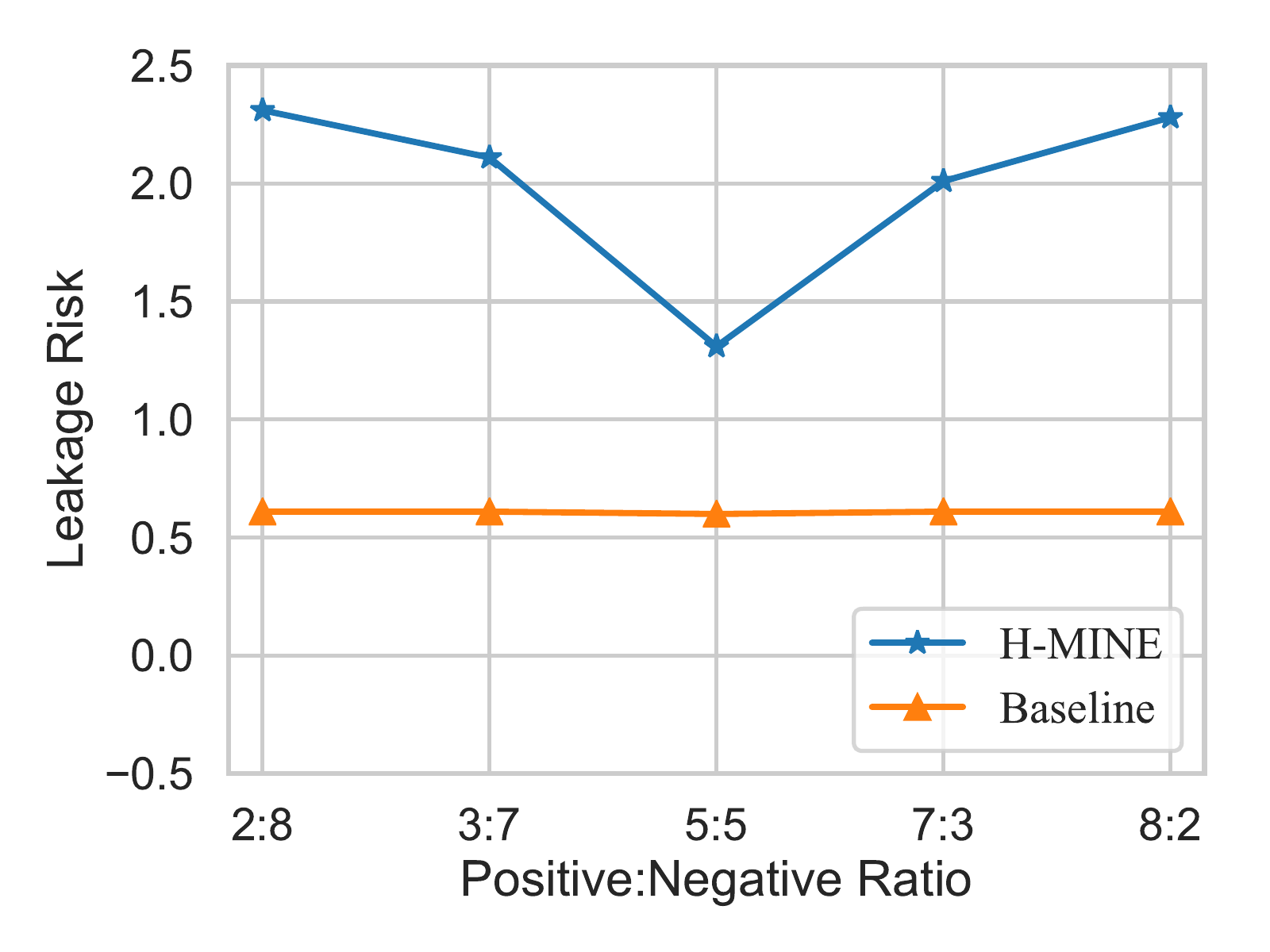}
  {\small (b) Leakage risk with unbalanced datasets, $epoch=3$, $B=3$.}
\end{minipage}
\caption{Inherent factor analysis.}
\label{fig:step_unbalance}

\end{figure}

Firstly, FedSGD is an iterative process.
Clients calculate their gradients over the evolving parameter values.
At different timesteps, the gradients may present different levels of information about the batched data, due to its interaction with the current model status.
Fig.\ref{fig:step_unbalance}(a) presents $I(X_B;G^t)$ verses the iteration step $t$.
The MI is the largest at the first epoch, while it gradually decreases over the optimization process.
At initial steps, every data point is new to the model.
Thus the gradients possess strong information about how the model misses to fit the observed data.
When the model starts to gain a reasonable form, the amount of miss-fit shrinks, therefore revealing less about the observed data.
As the task model converges, the MI between the gradients and the raw data also stablizes at a certain level.



Secondly, it is speculated that the data distribution itself also affects the risk of information leakage \cite{hsu2020federated}.
In this experiment, we simulate a client dataset with different ratios of positive and negative entries, and investigate the corresponding variation of $I(X_B;G^t)$.
As shown in Fig \ref{fig:step_unbalance}(b), the more unbalanced the dataset is, the more information about the batched data is preserved in the gradients.
The results verify that, gradients on unbalanced data distributions are more vulnerable to inference attacks.



\section{Conclusions}
In this paper, we proposed a novel security metric for FL clients based on high-dimensional mutual information estimation.
The proposed algorithm H-MINE can effectively and faithfully approximate the mutual information between model gradients and the original batched data.
Therefore, it provides a quantitative metric for potential risk alarms on the client side.
Analysis on the inherent factors of information leakage risk suggest data holders to be cautious with initial training steps and unbalanced data distributions.

\section{ACKNOWLEDGEMENTS}

This paper is supported by National Key Research and Development
Program of China under grant No. 2018YFB1003500,
No. 2018YFB0204400 and No. 2017YFB1401202. Corresponding
author is Jianzong Wang from Ping An Technology
(Shenzhen) Co., Ltd.




\vfill\pagebreak

\bibliographystyle{IEEEbib}
\bibliography{strings,refs}

\end{document}